\documentclass[twocolumn]{aastex631}
\pdfoutput=1
\usepackage{amsmath}
\usepackage{epsf}
\usepackage{color}
\usepackage{color}
\usepackage{enumitem}
\usepackage{mathtools}
\usepackage{savesym}
\savesymbol{tablenum}
\usepackage{siunitx}
\restoresymbol{SIX}{tablenum}
\usepackage{booktabs}

\usepackage{xcolor}
\usepackage{soul}

\usepackage{siunitx}
\usepackage{multirow}
\usepackage{makecell}

\usepackage[mathscr]{euscript}
 \let\mathscr\relax
\usepackage[scr]{rsfso}

\shorttitle{$z\sim3$ AGN from HETDEX Survey}
\shortauthors{Tardugno Poleo et al.}

\turnoffedit 

\begin{document}
\title{Identifying Active Galactic Nuclei at $z\sim3$ from the HETDEX Survey Using Machine Learning}

\author{Valentina Tardugno Poleo}
\affiliation{Department of Physics, The University of Texas at
  Austin, Austin, TX 78712, USA}
\email{vtardugno@utexas.edu}

\author[0000-0001-8519-1130]{Steven L. Finkelstein}
\affiliation{Department of Astronomy, The University of Texas at Austin, Austin, TX 78712, USA}

\author{Gene Leung}
\affiliation{Department of Astronomy, The University of Texas at Austin, Austin, TX 78712, USA}

\author[0000-0002-2307-0146]{Erin Mentuch Cooper}
\affiliation{Department of Astronomy, The University of Texas at Austin, Austin, TX 78712, USA}
\affiliation{McDonald Observatory, The University of Texas at Austin, Austin, TX 78712, USA}

\author[0000-0002-8433-8185]{Karl Gebhardt}
\affiliation{Department of Astronomy, The University of Texas at Austin, Austin, TX 78712, USA}

\author[0000-0003-2575-0652]{Daniel J. Farrow}
\affiliation{University Observatory, Fakult\"at f\"ur Physik, 
Ludwig-Maximilians University Munich, Scheinerstrasse 1, 81679 Munich, 
Germany}
\affiliation{Max-Planck Institut f\"ur extraterrestrische Physik, 
Giessenbachstrasse 1, 85748 Garching, Germany}

\author[0000-0003-1530-8713]{Eric Gawiser}
\affiliation{Department of Physics and Astronomy, Rutgers, the State University, Piscataway, NJ 08854, USA}

\author[0000-0003-2307-0629]{Greg Zeimann}
\affiliation{Hobby Eberly Telescope, The University of Texas at Austin, Austin, TX, 78712, USA}

\author[0000-0001-7240-7449]{Donald P. Schneider}
\affiliation{Department of Astronomy and Astrophysics, The Pennsylvania State University,
   University Park, PA 16802}
\affiliation{Institute for Gravitation and the Cosmos, The Pennsylvania State University,
   University Park, PA 16802}

\author{Leah Morabito}
\affiliation{Centre for Extragalactic Astronomy, Department of Physics, Durham University, Durham, DH1 3LE, UK}

\author{Daniel Mock}
\affiliation{Department of Physics, Florida State University, Tallahassee, FL, USA}

\author[0000-0001-5561-2010]{Chenxu Liu}
\affiliation{South-Western Institute for Astronomy Research, Yunnan University, Kunming, Yunnan, 650500, People’s Republic of China}
\affiliation{Department of Astronomy, The University of Texas at Austin, Austin, TX 78712, USA}

\begin{abstract}
We used data from the Hobby-Eberly Telescope Dark Energy Experiment (HETDEX) to study the incidence of AGN in continuum-selected galaxies 
at $z\sim3$. From optical and infrared imaging in the 24 deg$^{2}$ Spitzer HETDEX Exploratory Large Area (SHELA) survey, we constructed a sample of photometric-redshift selected $z\sim3$ galaxies. We extracted HETDEX spectra at the position of 716 of these sources and used machine learning methods to identify those which exhibited AGN-like features. The dimensionality of the spectra was reduced using an autoencoder, and the latent space was visualized through t-distributed stochastic neighbor embedding (t-SNE). Gaussian mixture models were employed to cluster the encoded data and a labeled dataset was used to label each cluster as either AGN, stars, high-redshift galaxies, or low-redshift galaxies. Our photometric redshift (photo-z) sample was labeled with an estimated 92\% overall accuracy, an AGN accuracy of 83\%, and an AGN contamination of 5\%. The number of identified AGN was used to measure an AGN fraction for different magnitude bins. The UV absolute magnitude where the AGN fraction reaches 50\% is $M_{UV} = -$23.8. 
When combined with results in the literature, our measurements of AGN fraction imply that the bright end of the galaxy luminosity function exhibits a power-law rather than exponential decline, with a relatively shallow faint-end slope for the $z \sim$ 3 AGN luminosity function.


\end{abstract}


\section{Introduction}\label{sec:intro}

The shape of the active galactic nuclei (AGN) rest-frame ultraviolet (UV) luminosity function, particularly the faint end, can provide insights into the potential contribution of AGN to the epoch of reionization \citep[e.g.,][]{Madau_2015,gia,Kulkarni_2019,Finkelstein_2019}. The luminosity function contains information about non-ionizing radiation emitted by AGN, which can be utilized to extrapolate the amount of ionizing photons emitted by such sources. A steepening slope with increasing redshift would suggest a potentially non-negligible contribution of AGN to the ionizing photon budget into the epoch of reionization, while a slope that becomes shallower (or stays fixed) with increasing redshift would fail to support such a hypothesis.

However, there are many uncertainties regarding the faint end of the AGN luminosity function, particularly at $z\geq3$, and thus, there is profuse interest in better constraining this function. One approach to make progress is to fit the combined AGN $+$ star-forming galaxy luminosity functions, now possible given a wealth of wide-field surveys \citep{stevans, adams, Harikane_2022, finkelstein22, zhang}.  For example, \citet{stevans} created a rest-frame UV $z = 4$ luminosity function of star-forming galaxies and AGN from the SHELA field but were unable to determine if the AGN luminosity function faint-end slope was shallow or steep. The shape of the faint end of the AGN luminosity function depends on the shape of the bright end of the galaxy luminosity function, where a power-law shaped bright end corresponds to a shallow AGN faint-end slope and an exponential (Schechter-like) decline corresponds to a steeper slope. An important parameter to determine the shape of the bright end and constrain the faint end of the luminosity function is the AGN fraction, i.e. the ratio of the number density of AGN to the total population at a given UV luminosity. Thus, calculating an AGN fraction in the luminosity range where AGN begin to overtake galaxies could aid in breaking the degeneracies in luminosity function fits.

Utilizing spectroscopic data from the Hobby-Eberly Telescope Dark Energy Experiment (HETDEX) \citep{hetdex}, and optical and infrared imaging in the $24$ deg$^2$ Spitzer HETDEX Exploratory Large Area (SHELA) survey \citep{shela}, we devised a method to measure the AGN fraction at $z\sim3$ using machine learning. In future research, our methodology could be applied to $z=4$ and beyond and help constrain the UV luminosity function of AGN. \hyperref[sec:data]{Section 2} of this paper focuses on the selection criteria for the photo-z, training, and validation samples. The dimensionality reduction of all the samples through an autoencoder neural network and t-Distributed Stochastic Neighbor Embedding (t-SNE) is described in detail in \hyperref[sec:reduction]{Section 3}. \hyperref[sec:clustering]{Section 4} discusses the clustering of the data and \hyperref[sec:agnfrac]{Section 5} shows our calculation of the AGN fraction. Lastly, \hyperref[sec:discussion]{Section 6} provides a discussion of the results, and \hyperref[sec:conclusion]{Section 7} concludes with a summary of the methods described in this paper and potential directions for future work. Throughout this paper, all magnitudes are provided in AB units \citep{abunits}. A 2013 Planck cosmology is assumed, where $H_0 = 67.8$ km s$^{-1}$ Mpc$^{-1}$, $\Omega_M = 0.307$, and $\Omega_\Lambda = 0.693$ \citep{planck}.

\section{Data}\label{sec:data}

\subsection{Photometric Redshift Sample}
Our sample of $z\sim3$ star-forming galaxies and AGN in the SHELA field was constructed from the photometric catalog of \citet{pdf} following the procedure described in \citet{stevans}, tailoring our criteria to select galaxies at $z \sim3$. We used imaging from the $u^\prime$, $g^\prime$, $r^\prime$, $i^\prime$, and $z^\prime$ optical bands from the Dark Energy Camera (DECam), the 3.6 $\mu$m and 4.5 $\mu$m mid-IR bands from Spitzer/the Infrared Array Camera (IRAC), and the near-IR $J$ and $K_s$ from VISTA-CFHT. We limited our sources' signal-to-noise ratio to be greater than or equal to 3.5 in the $r^\prime$ and $i^\prime$ photometric bands to reduce the incidence of spurious sources. 
The photometric redshift probability distribution functions (PDF) from \citet{pdf} were used to select sources around the desired redshift. The area under a given source's PDF at $z>1.5$ was required to be greater than 0.8, and the area under the PDF between $z = 2.5$ and $z = 3.5$ had to be greater than the area under the PDF for all other redshift bins of width 1, centered around integer values of $z$. This selection procedure produced 5388 potential $z \sim$ 3 sources in the SHELA field. As the primary feature used to select these galaxies is the Lyman break, this sample should be inclusive of both AGN and star-forming galaxies.

\subsection{HETDEX Spectra}

HETDEX is an unbiased spectroscopic survey 
collecting data at the 10-meter Hobby-Eberly Telescope (HET). 74 integral-field unit (IFU) fiber arrays installed at HET feed two low-resolution Visible Integral-field Replicable Unit Spectrographs (VIRUS) \citep{hill, hill2} that span a wavelength range of 3500-5500 \r{A}. The survey is set to cover the “Spring” field, extending over 390 deg$^2$, and the equatorial “Fall” field which covers 150 deg$^2$, for a total area of 540 deg$^2$ \citep{hetdex}. The SHELA field, a 24 deg$^2$ region of sky in the Sloan Digital Sky Survey (SDSS) \citep{york} Stripe 82 field \citep{shela}, was one of the fields targeted by HETDEX for repeat observations. The first HETDEX catalog (Mentuch Cooper et al., submitted) includes all observations up until late June, 2020, with over 240 thousand Lyman-alpha emitter candidates, and covers $\sim$10\% of the SHELA field.

Utilizing the celestial coordinates of the photometric-redshift selected sources and a search radius of 3$^{\prime\prime}$ for aperture, we extracted PSF-weighted HETDEX spectra at the sources' positions using HETDEX's customized python software \texttt{hetdex-api}\footnote{\url{https://github.com/HETDEX/hetdex_api/blob/master/hetdex_tools/get_spec.py}}. This resulted in a sample of 716 $z\sim3$ photometrically-selected sources with extracted HETDEX spectra. We then limited the spectra to wavelengths between 3645 and 5475 \r{A} to remove high noise regions near the spectral edges. The data were normalized by dividing the flux density values of each spectrum by that spectrum's maximum value, as this normalization yielded the best reconstructions from the autoencoder (see \S~\ref{sec:autoencoder}). Normalizing the data places all spectra on the same scale, a key pre-processing step in the machine learning pipeline. The described selection of HETDEX spectra is referred to throughout this paper as the photo-z sample. 

\subsection{Training and Validation Samples}
When training a neural network, training and validation samples are required. The training set is utilized by the network to learn the relationship of interest, while the validation set is used to assess the network's ability to generalize the relationship to include new data. To create our training and validation samples for our neural network, we collected HETDEX spectra from known stars, AGN, low-redshift ($z<0.5$), and high-redshift ($1.9 < z < 3.5$) star-forming galaxies. For the AGN, we selected a quasar training set from SDSS DR16 objects labeled as “Quasar” \citep{sdss}. The galaxies were drawn from the sample presented in \citet{adam}. We used the HETDEX star catalog from \citet{hawkins} to select stars that had a signal-to-noise ratio greater than 15 in the $r^\prime$ and $b^\prime$ photometric bands. Because the number of stars was orders of magnitude greater than the number of all other training objects, we selected every 90\textsuperscript{th} star to avoid an overrepresentation of stars in training. The data were split between training and validation sets in a 4:1 ratio, with 1,968 training sources and 490 validation sources. 22\% of the sources were stars, 23\% AGN, 42\% low-z galaxies, and 13\% high-z galaxies. As with the photo-z sample, both sets were limited to wavelengths between 3645 and 5475 \r{A}, and normalized by dividing the flux desnity values by the maximum flux density value.

\section{Dimensionality Reduction}\label{sec:reduction}

Each spectrum in our sample consisted of 914 flux values corresponding to wavelengths across the selected range. Analyzing datasets of high dimensionality, such as our photo-z sample, often presents challenges. Working with a high number of variables can affect the performance of certain machine learning algorithms. Moreover, storing and analyzing high-dimensional data can be a complication in the presence of limited storage space \citep{textbook}. Reducing the dimensions of our dataset allowed us to avoid the aforementioned complications. Projecting data onto lower dimensional spaces also serves as a data visualization tool. Thus, to make our sample more manageable and extract information more effectively, we utilized an autoencoder neural network to decrease the number of variables associated with each spectrum. To visualize the resulting encoding we employed t-SNE to project our data to a two dimensional space. Although t-SNE is a data-reduction tool in itself, reducing the data to manageable dimensions before employing t-SNE allows for the algorithm to better diminish noise and to decrease computation time \citep{tsne_doc}.

\subsection{Autoencoder Neural Network}\label{sec:autoencoder}

\begin{figure}[ht]
\centering
\includegraphics[width=0.45\textwidth]{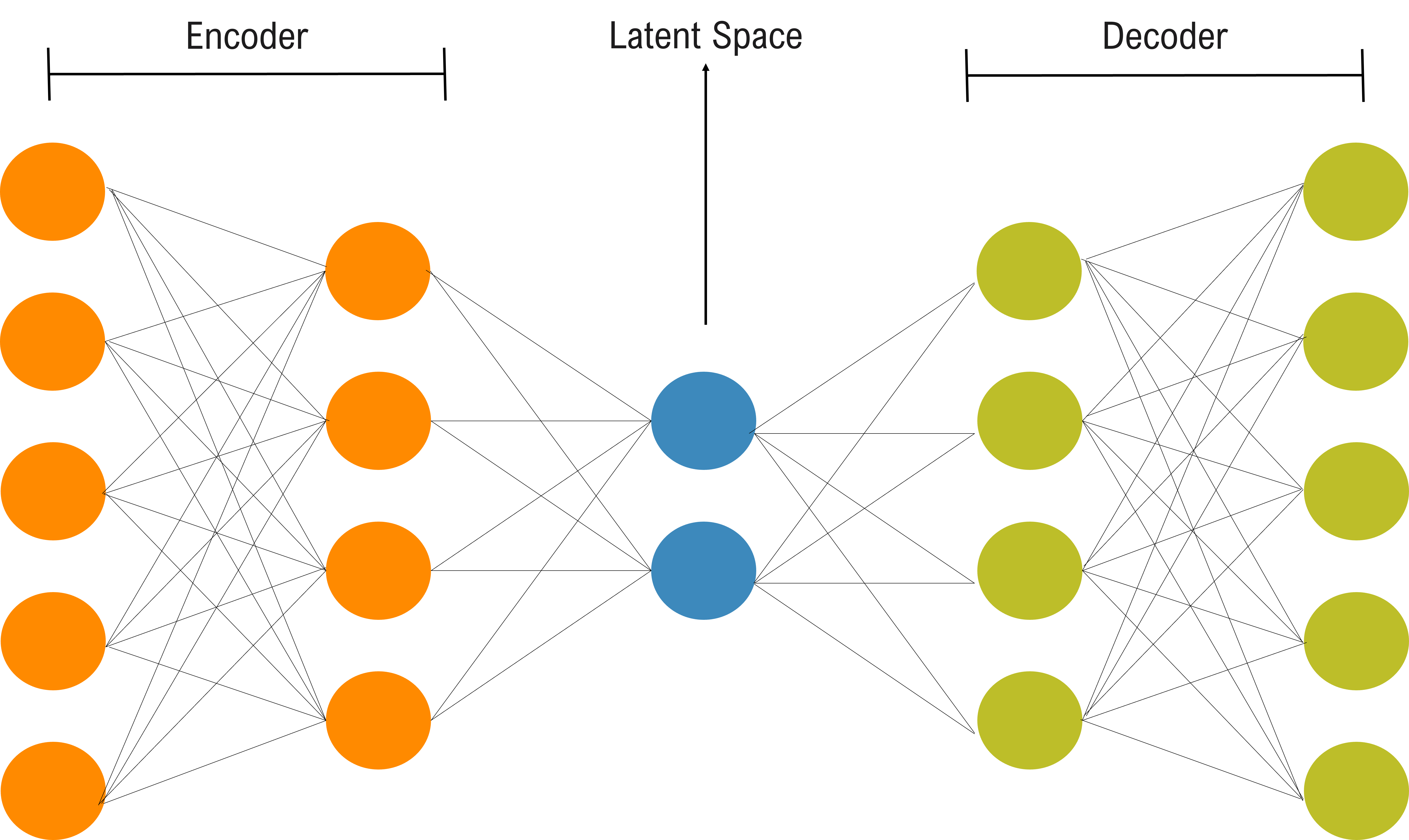} 
\caption{Schematic of an autoencoder neural network. The latent space contains a lower-dimensional representation of the network's inputs. We implemented an autoencoder to reduce the dimensionality of the spectra of our photo-z sample, resulting in a more manageable dataset that was later projected onto two dimensions through t-SNE and separated into clusters. This approach allowed for the removal of contaminants from our sample and a measurement of AGN fractions. }
\label{fig:autoenc} 
\end{figure}

To reduce the dimensionality of our spectra, we trained an autoencoder neural network using Keras \citep{keras} with the TensorFlow \citep{tf} backend. Autoencoders were first described in \citet{autoenc} as a neural network trained to output a reconstruction of the input. Autoencoders are composed of two networks, the encoder and the decoder (see Figure~\ref{fig:autoenc}). The encoder reduces the inputs' dimensions via matrix multiplication until it outputs vectors of the desired dimensions. The abstract space containing the encoding is referred to as the latent space. The decoder then takes as inputs the latent space representations and attempts, again via matrix multiplication, to recreate the original inputs. To learn more complex relationships beyond the linear nature of matrix multiplication, the networks feature activation layers, which apply non-linear functions to the hidden layers' nodes. 

The architecture of our autoencoder was built through hyperparameter optimization, i.e., selecting values for the parameters which control the network's learning in a way that improves the performance of the predictive model. The tuning parameters included the number of layers in the network, the number of nodes per layer, the optimization algorithm and its learning rate, dropout, and the type of activation function. As the hyperparameters were modified, the performance of the model was assessed by two measures. The first measure was the training loss, the error resulting from comparing the training input and its reconstruction. The network was designed to attempt to minimize the training error after every epoch. The second measure employed was the validation loss, which resulted from comparing the reconstruction of the validation set to the original validation spectra. Unlike the training loss, the validation loss was not used by the network to modify itself. In other words, the network was not learning from the validation data, but calculating how well the current configuration was reconstructing a previously unseen set.

\begin{figure}[h!]
\centering
\vspace{-4mm}
\includegraphics[width=0.5\textwidth]{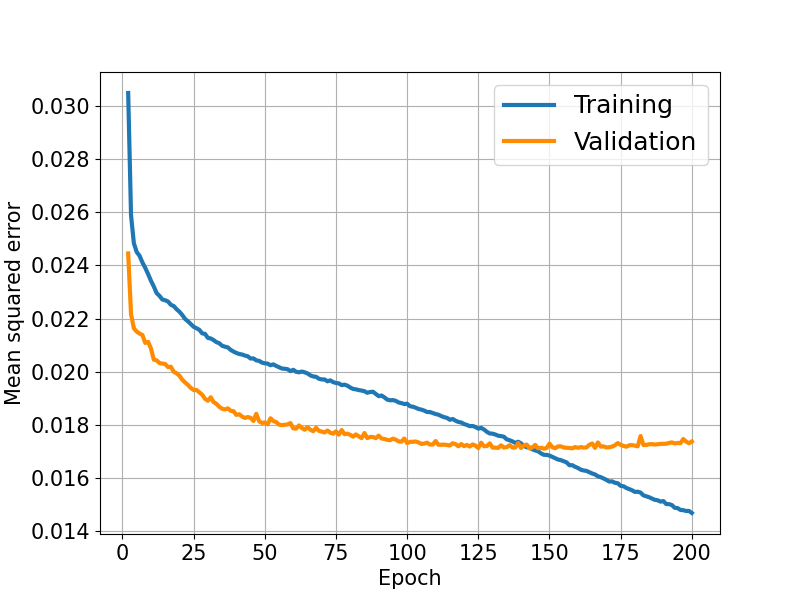} 
\caption{Training and validation loss curves. During training, we sought to minimize the training error while avoiding overfitting. Hence, the training stopped after 200 epochs on a decreasing training error and a plateauing validation error. }
\label{fig:loss} 
\end{figure}

To select the values of our hyperparameters, we sought to minimize both the training and validation losses (see Figure~\ref{fig:loss}). A decreasing training loss indicates that the model is learning patterns and relationships present in the training set. However, solely focusing on minimizing the training error can lead to overfitting, a network's failure to generalize to new or unseen data (e.g. \citet{textbook}). Hence, the validation error provides valuable information about the model's ability to effectively reconstruct unseen data.

After hyperparameter optimization, our autoencoder's encoder network was trained for 200 epochs and consisted of a 914-dimensional input layer, and one dense hidden layer with 436 nodes and a Sigmoid activation function, which assigned a value between 0 and 1 to the nodes' output. At every training epoch, a random 30\% of the hidden layer's nodes were dropped, resulting in a different configuration after every epoch. Dropout regularization allows the network to learn patterns in the spectra rather than memorize the training data. The resulting latent space was 30 dimensional. The decoder network followed a mirrored architecture, with an input layer of 30 dimensions, and a dense hidden layer with 436 nodes and a Sigmoid activation function. Dropout was omitted for the decoder network. The decoder's output layer was 914-dimensional, the same size as the encoder's input layer, and had no activation function. During training, the network utilized the Adam Optimizer \citep{adam_opt} with a learning rate of 0.0005 to minimize the mean-squared error of the decoder's reconstructed spectra when compared to the original input. 

After training the autoencoder, we utilized the encoder network to reduce the dimensionality of our training, validation, and photo-z-selected samples. To encode the photo-z sample, we inputted the spectra into the encoder, whose output was a 30-dimensional representation of the originally 914-dimensional spectra. The encoding carries the key features contained in the original spectra, in a much more compact format that allows for further analysis.

\begin{figure*}[t]
\centering
\includegraphics[width=0.5\textwidth]{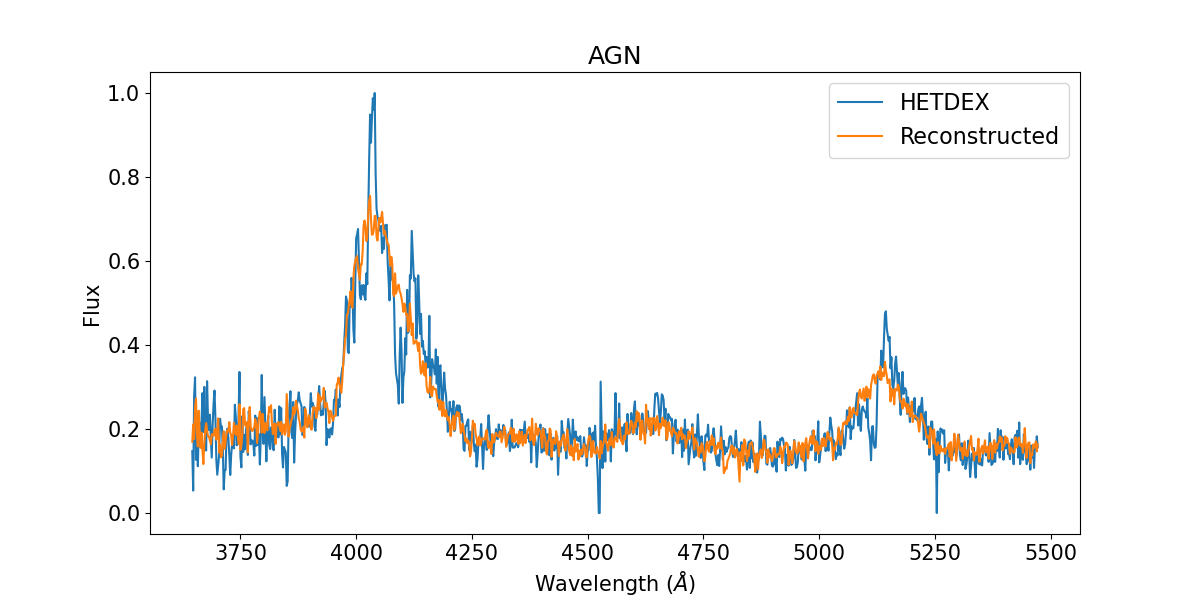}\hfill
\includegraphics[width=0.5\textwidth]{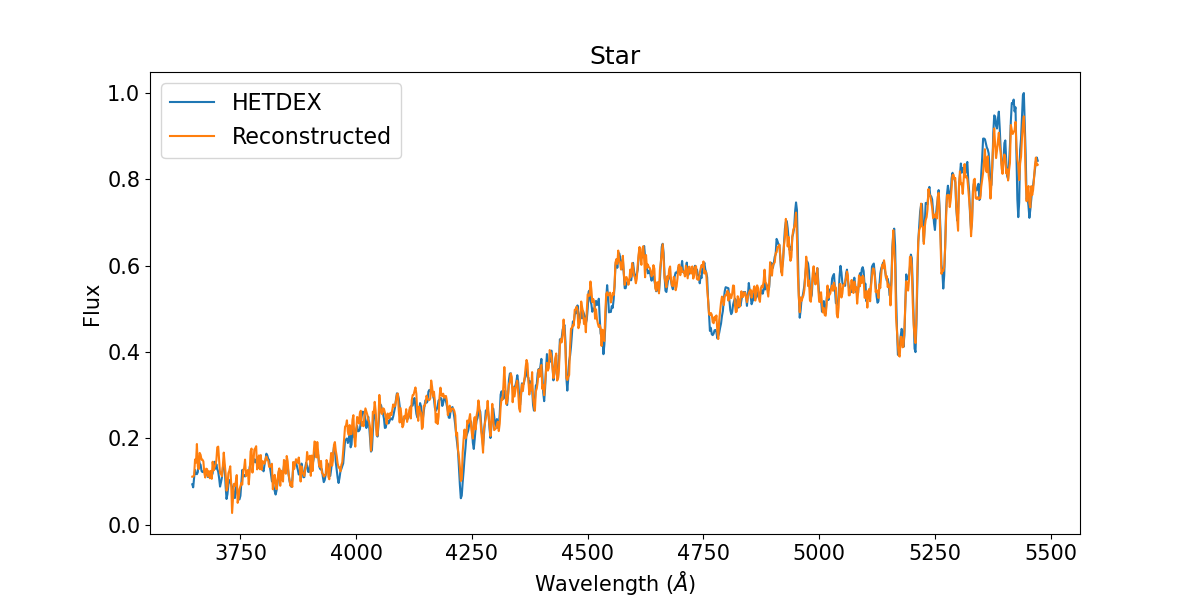}\hfill
\includegraphics[width=0.5\textwidth]{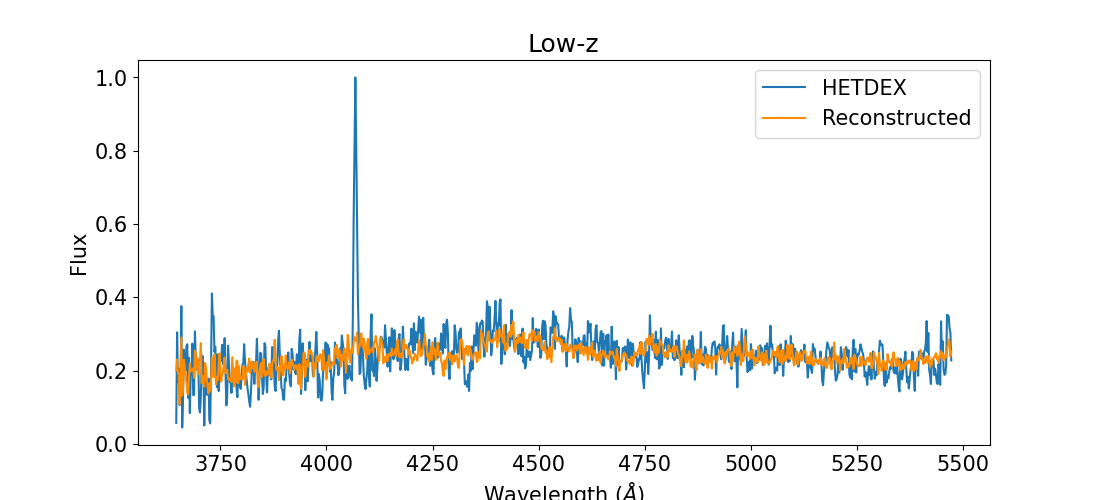}\hfill
\includegraphics[width=0.5\textwidth]{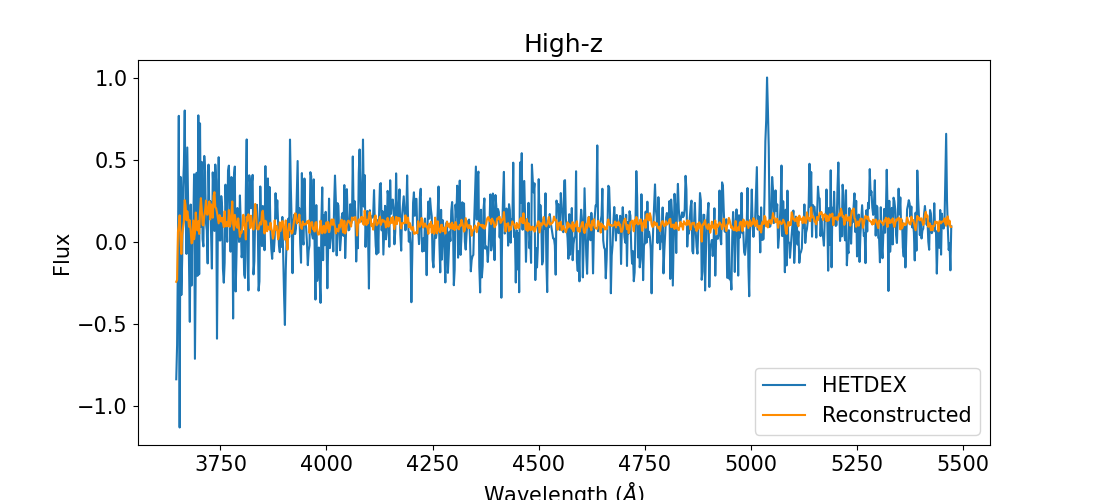}\hfill
\caption{Example spectra for an AGN (upper left), a star (upper right), a low redshift galaxy (bottom left), and a high redshift galaxy (bottom right). The blue spectra represent normalized data extracted from the HETDEX survey. The orange spectra represent reconstructed spectra obtained from running the 30-dimensional encoding through the decoder network. Comparing the HETDEX spectra to their corresponding reconstructed spectra allows for a visualization of how well the encoding preserves information. While the key features of the AGN and stars appeared to be preserved, narrow emission lines from galaxies were not reconstructed adequately. This made the separation between low and high redshift galaxies less prominent (see \S~\ref{sec:discussion}).}
\label{fig:recon} 
\end{figure*}

Upon visually inspecting the reconstructed spectra (see Figure~\ref{fig:recon}), we observed that the decoder adequately reconstructed spectral trends. The decoder was particularly effective in reconstructing spectral features such as broad emission lines, characteristic of AGN spectra, and stellar absorption lines. However, the reconstructions appear to be unable to capture narrow emission lines (see \S~\ref{sec:discussion}). Other spectral features such as noise levels, the presence of bright lines, and continuum can be exploited to distinguish between high and low-redshift galaxies.  

\subsection{t-SNE}

We employed a second dimensionality reduction algorithm, t-SNE \citep{tsne}, with the intention of visualizing the photo-z sample to identify differences between different astronomical objects' spectra. t-SNE maps higher-dimensional data to a two-dimensional space, allowing for the visualization of data in a simple two-dimensional display.

\begin{figure}[h]
\centering
\includegraphics[width=0.5\textwidth]{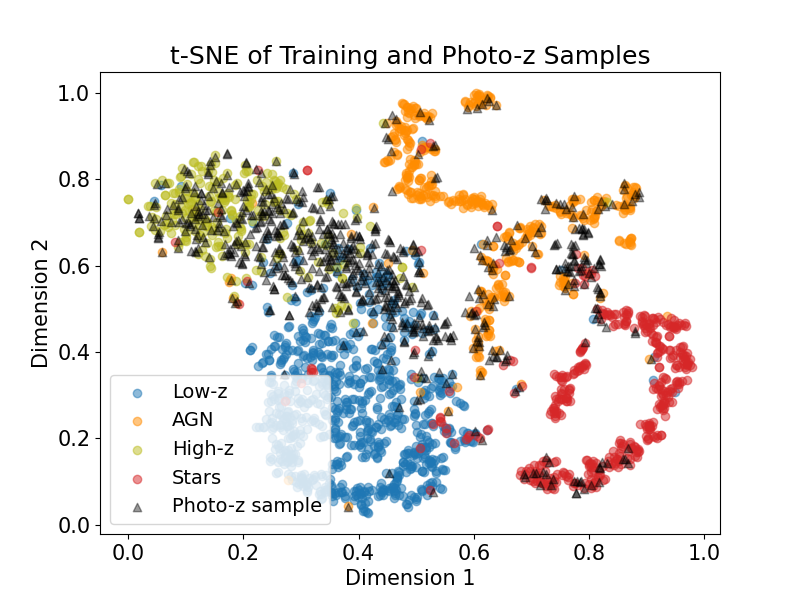} 
\caption{t-SNE plot of the combined training (colored circles) and photo-z (black triangles) samples. t-SNE allows for a visualization of our samples in a two-dimensional plane, revealing where each spectrum lies in relationship to the others. Astronomical objects form clusters with others of their kind, a feature that allowed us to employ a clustering algorithm to label our photo-z sample. Most of the spectra from the photo-z sample appeared to be concentrated around the high redshift galaxies and AGN from the training sample, which was to be expected based on our selection criteria.}
\label{fig:tsne} 
\end{figure}

After encoding all the samples, we combined the training and photo-z sets and employed t-SNE using the scikit-learn library (see Figure~\ref{fig:tsne}). The perplexity, which measures the effective number of neighbors \citep{tsne}, was set to 10 as it resulted in the most distinct separation of astronomical objects when visualizing the training and validation samples. Perplexity values should range between 5 and 50, with larger datasets generally requiring a larger value \citep{tsne_doc}. The maximum number of iterations was set to 3000. The resulting dataset contained a coordinate pair corresponding to each spectrum, which allowed for the creation of a two-dimensional plot to visualize the separation between different astronomical objects. 

\section{Clustering}\label{sec:clustering}

Figure~\ref{fig:tsne} reveals that each astronomical object appears closer to other spectra of their kind, resulting in four distinct clusters representing the four types of astronomical objects included in the training sample. To identify the clusters and therefore the astronomical objects based on their separation in the t-SNE diagram, we employed several different clustering algorithms, including Gaussian mixture models, agglomerative clustering, Density Based Spatial Clustering of Applications with Noise (DBSCAN), and spectral clustering, and compared their performance. Gaussian mixture models are ``parametric probability density function[s] represented as a weighted sum of Gaussian component densities" \citep{gauss}. Agglomerative clustering employs a bottom-up approach to hierarchical clustering, where all data points begin as their own cluster which are later merged together based on linkage distance \citep{AC}. DBSCAN operates under a ``density-based notion of clusters," and ``requires only one input parameter and supports the user in determining an appropriate value for it" \citep{dbscan}. Lastly, spectral clustering allows the user to ``apply clustering to a projection of the normalized Laplacian" \citep{SC}.

To identify the algorithm that best clustered the data, we applied them all to the t-SNE of the combined training and validation samples, leaving out the photo-z sample to avoid bias in selecting the clustering algorithm. From a simple visual inspection of the clustering on the t-SNE plot, the two algorithms that best clustered the training and validation sets were Gaussian mixture models and spectral clustering, with further evaluation needed to identify the best one. We calculated the total accuracy (overall percentage of correctly classified sources), AGN accuracy (percentage of true AGN that were predicted as AGN out of all true AGN), and contamination in AGN sample (percentage of non-AGN incorrectly labeled as AGN out of all the predicted AGN), using the validation set for both algorithms; the results are summarized in Table~\ref{tab:comp}. We determined that Gaussian mixture models performed better with the validation set and thus selected this algorithm to cluster our photo-z data. To further understand how the algorithm performed in the different groups of astronomical objects, a confusion matrix was calculated (see Table~\ref{tab:conf}). The confusion matrix entries contain the predicted and actual labels of a sample, which allows for the visualization and evaluation of an algorithm's classifying performance.

\begin{table}[ht!]
\begin{center}
\begin{tabular}{c c c} 
 \hline
   & Gaussian mixture & Spectral \\
 \hline
 Total accuracy & 92\% & 83\%  \\ 
 AGN accuracy & 83\% & 80\% \\ 
 AGN contamination & 5\% & 4\%\\
 \hline
\end{tabular}
\end{center}
\caption {Comparison between Gaussian mixture models and spectral clustering (see \S~\ref{sec:clustering}). All values were calculated using the validation data only. The total and AGN accuracy were the total percentage of correctly classified spectra and the percentage of correctly classified AGN respectively. The AGN contamination was the percentage of non-AGN incorrectly labeled as AGN out of all the predicted AGN. From comparing both clustering algorithms, we concluded that Gaussian mixture models better captured the clusters formed by the t-SNE of the training and validation samples. Thus, we chose the former algorithm to cluster the t-SNE of the training and photo-z sets in order to label the photo-z sample's spectra.} 
\label{tab:comp} 
\end{table}

\begin{table}[h]
\hskip-0.5cm\begin{tabular}{ccccc} \toprule & \multicolumn{4}{c}{Predicted labels} 
\\ \cmidrule{2-5} True labels & AGN & High-z & Low-z & Stars\\ \midrule 
AGN & 92 & 5 & 11 & 3\\ 
High-z &1 &60 &2 &0\\ 
Low-z & 1& 11& 197&0\\ 
Stars &3 & 1 & 3&100\\ 
\bottomrule 
\end{tabular} 
\caption {Confusion matrix of the validation data. The rows contain the actual labels and the columns contain the labels predicted by employing Gaussian mixture models on the t-SNE of the training and validation sets. The diagonal represents all correctly labeled sources. Through our methodology we were able to label the spectra from our photo-z sample in order to remove contaminants and find the number of AGN and high-z galaxies present in our sample. The AGN and high-z counts allowed us to calculate AGN fractions.}
\label{tab:conf} 
\end{table}

\begin{figure}[h]
\centering
\includegraphics[width=0.5\textwidth]{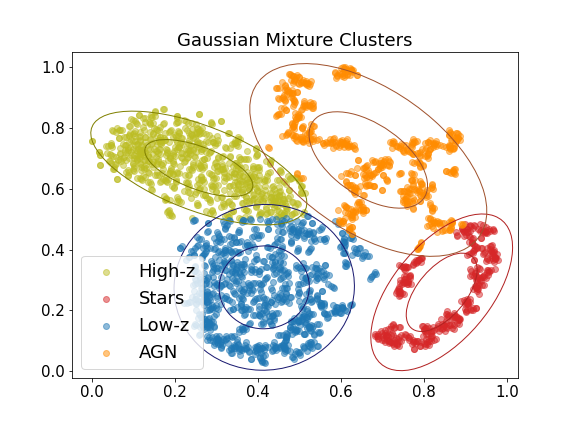} 
\caption{Clusters identified through Gaussian mixture models on the t-SNE of the combined training and photo-z samples. The four pairs of ellipses represent two standard deviations away from the mean of each cluster, as estimated by the sklearn.mixture.GaussianMixture class. The clustering algorithm appears to adequately capture the clusters formed by each group of astronomical objects. Clustering allowed us to classify the photo-z spectra and therefore remove contaminants and find an AGN fraction.}
\label{fig:cluster} 
\end{figure}

We used the sklearn.mixture python package to cluster the t-SNE of the combined training and photo-z samples, implementing Gaussian mixture models with k-means as the initialization method and four mixture components (one for each astronomical object type in the training sample). The resulting clusters are displayed in Figure~\ref{fig:cluster}. Using the training sample, we assigned a label to each cluster. We then assigned a label to each of the photo-z sample spectra, depending on which cluster they belonged to. Out of the 716 spectra, 147 were labeled AGN, 438 high-redshift galaxies, 100 low-redshift galaxies, and 31 stars.

\section{AGN Fraction}\label{sec:agnfrac}

\begin{figure}[t]
\centering
\includegraphics[width=0.5\textwidth]{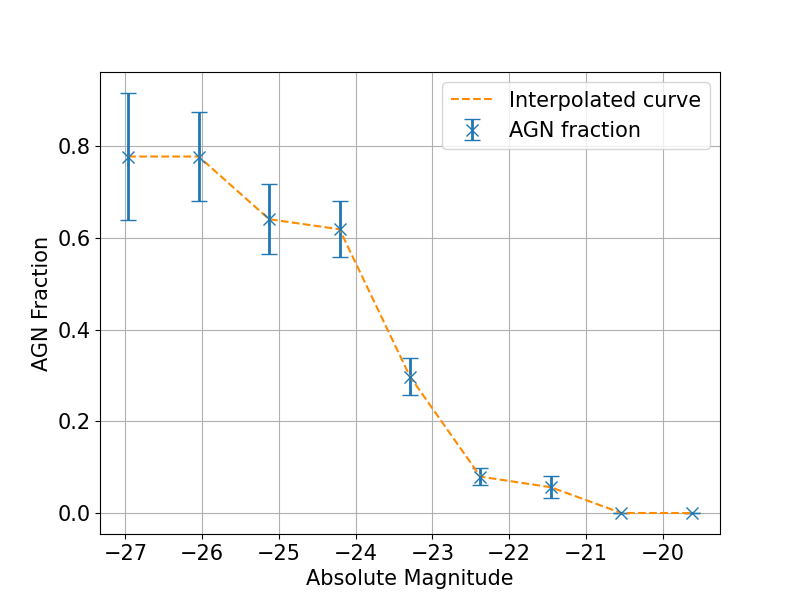} 
\caption{AGN fraction as a function of $r$-band absolute magnitude and an interpolated curve. The two brightest bins have an AGN fraction of $0.8 \pm 0.1$. The AGN fraction drops to $0.37 \pm 0.03$ at a magnitude of $-23.5$, and approaches zero at the faintest bins. Our calculated AGN fractions imply a shallower faint-end slope for the AGN luminosity function at redshift three.}
\label{fig:agnfrac} 
\end{figure}

Having labeled all the photo-z sample spectra, we determined the AGN fraction at different magnitude bins. We first removed the sources that were labeled as stars and low-z, leaving the identified high-z galaxies and AGN. Using the $r$-band flux of the remaining sources, which corresponds to rest-frame $\lambda \sim$ 1500-2000 \r{A} across our redshift range, we calculated their absolute magnitudes by applying the cosmological distance modulus at the spectroscopic redshift and separated them into nine magnitude bins. We then found the AGN fraction for each of the magnitude bins by finding the ratio of AGN to the sum of AGN and high-z galaxies (see Figure~\ref{fig:agnfrac}). We assumed that the uncertainties for the AGN and high-z galaxy counts were consistent with a Poisson distribution. If $x$ was the number of counts, the uncertainty $\sigma_x$ was set to $\sqrt{x}$. The uncertainties were then propagated to find the uncertainty in the AGN fraction. 

The $z \sim$ 3 AGN fraction of 50\% occurs at a UV absolute magnitude of $-$23.8, which is consistent with the \citet{stevans} results where the bright end of the galaxy luminosity function follows a power-law decline and the faint end of the AGN luminosity function is consistent with a shallower slope. At a UV magnitude of $-$23.5, \citet{stevans} predicts an AGN fraction of $\sim18$\% for a double-power-law fit to the galaxy luminosity function, and $\sim94$\% for a Schechter fit. At this magnitude, our measured AGN fraction was $37 \pm 3\%$, which is more closely related to the predictions for the power-law fit. Through both measurements, our results suggest an AGN fraction more consistent with a double-power law shape for the star-forming galaxy luminosity function, and thus a shallower AGN faint end slope. Due to the wavelength restriction of HETDEX, our measurements are at $z \sim$ 3, while the \citet{stevans} luminosity function is at $z =$ 4, thus future work with redder spectra can explore whether our results hold at this slightly larger redshift. 

We note that our AGN fraction does not reach unity, even at $M < -$ 26.  As it is highly unexpected to find star-forming galaxies at these luminosities, we visually inspected the spectra classified as high-redshift galaxies in these bins. The spectra appeared noisy 
and did not exhibit any visible high-z or AGN features. We conclude that the likely explanation for these sources is that their photometric redshifts have been incorrectly estimated, a factor that can soon be improved with the new SHELA photometric catalog (Leung et al.\ in prep) which incorporates new HSC imaging that is at least one magnitude deeper than the DECam imaging in the current catalog. 

\section{Discussion}\label{sec:discussion}
\subsection{Autoencoder}
The autoencoder approach to reduce the dimensionality of the spectra had both benefits and shortcomings. The reconstructed spectra effectively captured broad emission lines, which are generally present in AGN spectra. The autoencoder also succeeded at reconstructing stellar spectra. However, the autoencoder generally failed to reconstruct narrow emission lines and thus separating high and low-redshift galaxies became a challenge. Although the network may have used other features such as continuum to differentiate between galaxies, improving the autoencoder's ability to recognize narrow emission lines could significantly improve the separation between low and high-redshift galaxies. Moreover, the described methodology did not allow for sources to be labeled as noise. Therefore, pure noise spectra could have been labeled as high-redshift galaxies, which may impact the calculated AGN fractions (see \S~\ref{sec:conclusion}).

\subsection{Gaussian Mixture Models}

Gaussian mixture models significantly outperformed agglomerative clustering, spectral clustering, and DBSCAN. Upon visual inspection, it is not evident that the data is composed of a mixture of Gaussian distributions. However, as exemplified by our results, Gaussian mixture models can successfully cluster data that does not appear to follow a Gaussian distribution. Modeling our data using a mixture of Gaussian probability distributions, despite its potential non-Gaussian nature, resulted in almost all of the data points in each cluster falling within two standard deviations of the estimated mean (see Figure~\ref{fig:cluster}).

\subsection{Implications and Further Work}
The described methodology serves as a proof-of-concept that, if applied to other redshifts, could better constrain the faint end of the AGN UV luminosity function. In particular, finding the AGN fractions at $z\sim4$ could break the degeneracy of the faint-end slope and identify whether the shape of the AGN luminosity function presented by \citet{stevans} is best described by a shallow or a steep faint-end slope. Combining our results with studies at other redshifts, we can explore if there is a steepening faint-end slope with increasing redshift, which would imply a potential contribution by faint AGN to the ionizing photon budget at the end of reionization. However, our results at $z\sim3$ are consistent with a shallower faint-end slope of the AGN luminosity function. If similar results were found at higher redshifts, the findings may suggest a smaller AGN contribution. The faint-end slope remains a key parameter to investigate the role that AGN played in reionizing the intergalactic medium.

\section{Conclusions}\label{sec:conclusion}

In this paper, we develop a method to measure the AGN fraction in a photometrically selected sample at $z\sim3$ using machine learning. We used optical and infrared imaging from the SHELA field to select potential AGN and star-forming galaxies, and extracted spectroscopic data from HETDEX at these sources' positions. To reduce the dimensionality of the resulting 716 spectra, we employed the encoder network of an autoencoder. We used t-SNE to visualize the encoded data and Gaussian mixture models to identify clusters. Using the labels of the training data we assigned a label to each cluster and thus to each spectrum in our photo-z sample, allowing us to remove stars and low-redshift galaxies and to calculate an AGN fraction. 

When applying the described methodology to a validation set, we labeled the spectra with an accuracy of 92\% and 5\% AGN contamination. Hence, we were able to apply these methods to our unlabeled photo-z sample from the SHELA field to measure an AGN fraction. Our method resulted in 147 sources being classified as AGN and 438 as high-redshift galaxies, which yielded an AGN fraction of 50\% at a UV absolute magnitude of -23.8. This fraction can be used to define the shape of the faint end of the UV luminosity function and assess the contribution of faint AGN to reionization. If this result is similar at $z =$ 4, it would break the luminosity function degeneracy found by \citet{stevans} in favor of a shallower AGN faint end slope, and imply a smaller contribution from AGNs to the ionizing photon budget at higher redshift. 

However, there are changes that could be implemented that may increase confidence in the results. For instance, the influence of noise on the described methodology merits a thorough analysis. Including pure noise spectra in the training and validation sets may help avoid the mislabeling of noise in the photo-z sample as high redshift galaxies that could be lowering the measured AGN fraction. However, including noise as a category may also result in high-z galaxies with no distinguishable emission line being misclassified as noise and therefore an artificially higher AGN fraction. The methods chosen to address noise in the analysis have the potential to influence the AGN fraction and hence are worth exploring in future work. 

Further research is needed to fulfill our motivation of constraining the faint-end slope of the AGN UV luminosity function at $z=4$ and beyond to assess AGN contribution to the ionizing photon budget. Future studies could focus on constructing a $z\sim3$ luminosity function from our measured AGN fractions. Applying our methodology to analyze similar spectra with different photometric redshifts or including sources outside of the SHELA field may also be of interest.  Moreover, the described methods could be applied to $z=4$ data to find an AGN fraction and break the degeneracy identified in \citet{stevans}. Lastly, following the same procedure at higher redshifts could allow for a study of the evolution of the faint-end slope of the luminosity function, which may provide insights into the role that AGN played, if any, during the epoch of reionization.

\section{Acknowledgements}\label{sec:acknowledge}

We acknowledge that the location where most of this work took place, the University of Texas at Austin, sits on indigenous land. The Tonkawa lived in central Texas and the Comanche and Apache moved through this area. We pay our
respects to all the American Indian and Indigenous Peoples and communities who have been or have become a part of these lands and territories in Texas, on this piece of Turtle Island.

HETDEX is led by the University of Texas at Austin McDonald Observatory and Department of Astronomy with participation from the Ludwig-Maximilians-Universit\"{a}t M\"{u}nchen, Max-Planck-Institut f\"{u}r Extraterrestrische Physik (MPE), Leibniz-Institut f\"ur Astrophysik Potsdam (AIP), Texas A\&M University, The Pennsylvania State University, Institut f\"ur Astrophysik G\"ottingen, The University of Oxford, Max-Planck-Institut f\"ur Astrophysik (MPA), The University of Tokyo, and Missouri University of Science and Technology. In addition to Institutional support, HETDEX is funded by the National Science Foundation (grant AST-0926815), the State of Texas, the US Air Force (AFRL FA9451-04-2-0355), and generous support from private individuals and foundations.

The authors acknowledge the Texas Advanced Computing Center (TACC) at The University of Texas at Austin for providing high performance computing, visualization, and storage resources that have contributed to the research results reported within this paper. URL: http://www.tacc.utexas.edu

VTP, SLF and GL acknowledge support from the National Science Foundation through grant AST-1908817.  The observations were obtained with the Hobby Eberly Telescope (HET), which is a joint project of the University of Texas at Austin, the Pennsylvania State University, Ludwig-Maximilians-Universit{\"a}t M{\"u}nchen and Georg-August-Universit{\"a}t G{\"o}ttingen. The HET is named in honor of its principal benefactors, William P. Hobby and Robert E. Eberly

VIRUS is a joint project of the University of Texas at Austin, Leibniz-Institut f{\"u}r Astrophysik Potsdam (AIP), Texas A\&M University (TAMU), Max-Planck Institut f{\"u}r Extraterrestrische Physik (MPE), Ludwig Maximilians-Univers{\"a}t M{\"u}nchen, Pennsylvania State University, Institut f{\"u}r Astrophysik G{\"o}ttingen, University of Oxford, and the Max-Planck-Institut f{\"u}r Astrophysik (MPA).

The authors acknowledge the Texas Advanced Computing Center (TACC) at The University of Texas at Austin for providing high performance computing, visualization, and storage resources that have contributed to the research results reported within this paper

\bibliography{references}{}
\bibliographystyle{aasjournal}


\end{document}